\documentclass[prd,showpcs,amsmath,amssymb,nofootinbib,longbibliography,twocolumn,showpacs,notitlepage,superscriptaddress]{revtex4-1}
\usepackage{multirow}
\usepackage{epsfig}
\usepackage{amsmath}
\usepackage{bm}
\usepackage{times}
\usepackage{graphicx}
\usepackage{color}
\usepackage{slashed}
\usepackage{xcolor}
\usepackage{graphicx}
\usepackage{amsmath}
\usepackage[latin1]{inputenc}
\usepackage{hyperref}
\usepackage{soul}
\usepackage{lipsum}

\def\bea{\begin{eqnarray}}
\def\eea{\end{eqnarray}}
\def\bean{\begin{equation*}}
\def\eean{\end{equation*}} 

\def\beaal{\begin{align}}
\def\eeaal{\end{align}}
\definecolor{OliveGreen}{rgb}{0,0.6,0}
\begin{document}

\begin{flushright}
KCL-PH-TH-2022-43
\end{flushright}

\title{Probing Early Universe Supercooled Phase Transitions with Gravitational Wave Data}
\author{Charles Badger}
\affiliation{Theoretical Particle Physics and Cosmology Group,  Physics Department, \\ King's College London, University of London, Strand, London WC2R 2LS, United Kingdom}
\author{Bartosz Fornal}
\affiliation{Department of Chemistry and Physics, Barry University, Miami Shores, Florida 33161, USA}
\author{Katarina Martinovic}
\affiliation{Theoretical Particle Physics and Cosmology Group,  Physics Department, \\ King's College London, University of London, Strand, London WC2R 2LS, United Kingdom}
\author{Alba Romero}
\affiliation{Institut de F\'\i sica  d'Altes Energies (IFAE), Barcelona Institute of Science and Technology, E-08193 Barcelona, Spain}
\author{Kevin Turbang}
\affiliation{Theoretische Natuurkunde, Vrije Universiteit Brussel, Pleinlaan 2, B-1050 Brussels, Belgium}
\affiliation{Universiteit Antwerpen, Prinsstraat 13, B-2000 Antwerpen, Belgium}
\author{Huai-Ke Guo}
\affiliation{Department of Physics and Astronomy, University of Utah, Salt Lake City, Utah 84112, USA}
\author{Alberto Mariotti}
\affiliation{Theoretische Natuurkunde and IIHE/ELEM, Vrije Universiteit Brussel, \& The  International Solvay Institutes, Pleinlaan 2, B-1050 Brussels, Belgium}
\author{Mairi Sakellariadou}
\affiliation{Theoretical Particle Physics and Cosmology Group,  Physics Department, \\ King's College London, University of London, Strand, London WC2R 2LS, United Kingdom}
\affiliation{Theoretische Natuurkunde and IIHE/ELEM, Vrije Universiteit Brussel, \& The  International Solvay Institutes, Pleinlaan 2, B-1050 Brussels, Belgium}
\author{Alexander Sevrin}
\affiliation{Theoretische Natuurkunde and IIHE/ELEM, Vrije Universiteit Brussel, \& The  International Solvay Institutes, Pleinlaan 2, B-1050 Brussels, Belgium}
\affiliation{Universiteit Antwerpen, Prinsstraat 13, B-2000 Antwerpen, Belgium}
\author{Feng-Wei Yang}
\affiliation{Department of Physics and Astronomy, University of Utah, Salt Lake City, Utah 84112, USA}
\author{Yue Zhao}
\affiliation{Department of Physics and Astronomy, University of Utah, Salt Lake City, Utah 84112, USA} 

\date{\today}

\begin{abstract}
We investigate the reach of the LIGO/Virgo/KAGRA detectors in the search for signatures of first-order phase transitions in the early Universe. Utilising data from the first three observing runs, we derive constraints on the parameters of the underlying gravitational-wave background, focusing on transitions characterised by strong supercooling. As an application of our analysis, we determine bounds on the parameter space of two representative particle physics models.
We also comment on the expected reach of third-generation detectors in probing supercooled phase transitions.
\end{abstract}
\vspace{-2mm}

\maketitle

\section{Introduction}
The 2015 discovery of gravitational waves (GWs) by the LIGO/Virgo collaboration, based on the data obtained at the twin LIGO detectors \cite{LIGOScientific:2016aoc} gave rise to the field of GW astronomy. Since then,  as many as $\mathcal{O}$(100) GW signals have been recorded \cite{LIGOScientific:2021usb,LIGOScientific:2021djp}. Those events include binary black hole mergers, binary neutron star mergers, and a black hole-neutron star merger. Apart from such individually detectable events, a GW background is also expected to be discovered with increased detector sensitivity. One contribution to this  background arises from the superposition of unresolved astrophysical sources \cite{Christensen:2018iqi}.
However, a more intriguing possibility is a contribution of cosmological origin. Several processes would give rise to such a cosmological GW background, including first order phase transitions (FOPTs) in the early Universe \cite{Kosowsky:1991ua}, inflation \cite{Turner:1996ck}, or topological defects such as cosmic strings \cite{Vachaspati:1984gt,Sakellariadou:1990ne} and domain walls \cite{Hiramatsu:2013qaa}. 
In this work, we concentrate  on the expected signatures from FOPTs. 

Although the particle content of the Standard Model (SM) alone is not sufficient for a FOPT to occur in the early Universe, 
FOPTs are a generic feature in a number of theories beyond the SM. Some examples include models with new physics at the electroweak scale  \cite{Grojean:2006bp,Vaskonen:2016yiu,Dorsch:2016nrg,Chala:2018ari,Alves:2018jsw},
hidden sectors \cite{Schwaller:2015tja,Breitbach:2018ddu,Croon:2018erz,Hall:2019ank}, 
dark matter
\cite{Hambye:2018qjv,Baldes:2018emh,Baldes:2021aph,Azatov:2021ifm,Baldes:2022oev},
unification \cite{Croon:2018kqn,Brdar:2019fur,Huang:2020bbe,Okada:2020vvb}, confinement \cite{Helmboldt:2019pan,Croon:2019ugf,Huang:2020crf}, 
baryon and/or lepton number violation \cite{Cheung:2012im,Katz:2016adq,Hasegawa:2019amx,Fornal:2020esl,Baldes:2021vyz,Azatov:2021irb}), neutrino mass models \cite{Brdar:2018num,Okada:2018xdh,DiBari:2021dri,Zhou:2022mlz}, 
axions \cite{Dev:2019njv,VonHarling:2019rgb,DelleRose:2019pgi}, supersymmetry breaking \cite{Demidov:2017lzf,Craig:2020jfv,Fornal:2021ovz}, or 
theories explaining flavour anomalies \cite{Greljo:2019xan,Fornal:2020ngq}.
[For a more complete list of references on models exhibiting FOPTs, we refer the reader to \cite{Caldwell:2022qsj}.]

This interplay between particle physics and GWs provides a unique opportunity to explore regions of parameter space otherwise  unreachable in typical particle physics experiments. 
Indeed, the new physics energy scales $\gtrsim 100 \ {\rm TeV}$ fall outside the range probed by Earth-based accelerators. However, precisely such large energy scales  can give  rise to a signal within the frequency range of the LIGO/Virgo/KAGRA (LVK)  detectors, since the peak frequency is expected to fall within the range $\sim (10-1000)$ Hz.

A particularly interesting scenario is when the FOPT is supercooled, which often increases the duration of the FOPT,  leading to an enhancement of the
GW signal. 
A prolonged period of supercooling can arise in theories with
Coleman-Weinberg-type symmetry breaking \cite{PhysRevD.7.1888} or
in strongly-coupled scenarios. 
Some models of this type are discussed in \cite{Creminelli:2001th,Randall:2006py,Nardini:2007me,Konstandin:2011dr,Jinno:2016knw,vonHarling:2017yew,Baratella:2018pxi,Prokopec:2018tnq,Marzo:2018nov,Ellis:2019oqb,DelleRose:2019pgi,Jinno:2019jhi,Lewicki:2020azd,Agashe:2019lhy,VonHarling:2019rgb,Ellis:2020awk,Ellis:2020nnr,Agashe:2020lfz,Lewicki:2021xku}. In what follows we apply our analysis of the LVK data to the theoretically  well-motivated  supercooled models described in \cite{Ellis:2019oqb,DelleRose:2019pgi}, and  derive the corresponding constraints on their parameter space. 

This is the first time the LVK data from the first three observing runs (O1, O2 and O3) is being used to set limits on the parameters of particle physics models through a FOPT search. 
So far, only general constraints on the GW background from FOPTs have been derived \cite{Romero_2021}. In particular, in the current analysis we apply our priors directly at the level of the particle physics parameters, e.g. particle masses and couplings.
This presents a  novel way of bridging the gap between data analysis and theoretical particle physics model building. %, thereby increasing reusability of the constraints provided by  GW data. 

The rest of the paper is organised as follows: In Sec.\,\ref{Sec:GW}, we review the expected GW spectra  from FOPTs, focusing  on the case of supercooling. In Sec.\,\ref{Sec:General constraints}, we place constraints on the GW spectra from supercooled FOPTs using LVK data. In Sec.\,\ref{Sec:Models}, we apply these constraints to two particle physics models that exhibit supercooling. Then, in Sec.\,\ref{Sec:PI}, we compare two different methods of analysing  the detectability of the GW background, namely the one utilising the power-law integrated sensitivity curves, and  the more intricate Bayesian data analysis. Finally, in Sec.\ref{Sec:3g}, we present an outlook on the reach of third-generation (3G) GW detectors.

\section{Gravitational waves  from supercooled phase transitions}
\label{Sec:GW}

The GW background is described in terms of its energy density spectrum via
\begin{equation}
    \Omega_{\rm GW}(f)=\frac{1}{\rho_{\rm c} }\frac{d\rho_{\rm GW}}{d\ln f} \ ,
\end{equation}
where $\rho_{\rm c}=3c^2H_0^2/(8\pi G)$ is the critical energy density of the Universe. This energy density depends on the parameters describing the FOPT, and therefore, on the shape of the effective potential, dictated by the parameters of the particle physics model and the temperature at which the transition occurs.

As the temperature of the Universe decreases, a new (true) vacuum with a lower energy density may appear, along with a potential barrier separating it from the high-temperature (false) vacuum. The transition between the two states corresponds to the formation of bubbles of true vacuum in various patches of the Universe, and their subsequent expansion.
The nucleation rate per unit volume of such bubbles can be roughly estimated as \cite{LINDE1983421}
\bea\label{gamma}
\Gamma(T) \,\sim \, T^4 \exp\left(-\frac{S(T)}{T}\right) \, ,
\eea
where $S(T)$ is the Euclidean action evaluated on the bubble solution interpolating between the false and true vacuum.
The onset of a FOPT occurs at the nucleation temperature $T_n$ at which $\Gamma(T_n) \approx H_n^4$, with $H_n$ denoting the Hubble parameter at that time, $H_n=H(T_n)$.\footnote{In the case of supercooled FOPTs, special care needs to be taken to make sure that bubble percolation is possible despite the exponential expansion of the false vacuum \cite{Ellis:2020nnr}. This has been verified to be true in the model parameter space we are considering.}

A FOPT can be described by four parameters: the bubble wall velocity $v_w$, the nucleation temperature $T_n$,  the inverse of the transition's duration in Hubble units $\beta/H_n$,
\bea
\frac{\beta}{H_n}  = T_n \frac{d}{dT}\left(\frac{S(T)}{T}\right)\bigg|_{T=T_n} \ ,
\eea
and the strength of the transition $\alpha$,
\bea
\alpha = \frac{\rho_{\rm vac}(T_n)}{\rho_{\rm rad}(T_n)}\ ,
\eea
which is the ratio
of the vacuum energy density to the radiation energy density at nucleation temperature.
Out of the four parameters $v_w$, $T_n$, $\beta/H_n$, $\alpha$, only the bubble wall velocity does not depend on the shape of the effective potential, and we will set it to $v_w = c$ (for a detailed discussion of the possible choices see \cite{Espinosa:2010hh,Caprini:2015zlo}). We note that the temperature of the thermal bath at the time when the GWs are produced is not $T_n$,  but rather the reheating temperature $T_{\rm RH}$, approximately given by
\begin{equation}
    T_{\rm RH}^4\simeq\frac{30}{\pi^2g_*}\Delta V,
\end{equation}
where $\Delta V$ is the potential difference between the true and false vacuum, and $g_*$ is the number of relativistic degrees of freedom which we fix to $g_*=100$ throughout the analysis. It is often the case that $T_{\rm RH}\gg T_n$, especially for supercooled phase transitions.
However,  for 
sufficiently fast reheating one has  $H(T_n)\simeq
H(T_{\rm RH})$, which implies that  $\beta/H_n\simeq\beta/H_{\rm RH}$  \cite{Caprini:2015zlo}. 

The phenomenon of supercooling occurs when the nucleation temperature is much lower than the scale of the symmetry breaking triggering the FOPT, leading to a large  FOPT strength,  $\alpha \gg 1$. Given the current sensitivity of LVK detectors, one may expect that 
FOPT GW signals in reach of the experiment would most probably come from a supercooled FOPT.

Several  processes  contribute to the GW signal from a FOPT. Those include  bubble collisions, sound waves, and turbulence, the last of which  will not be considered here, as it is generally subdominant  \cite{Kamionkowski:1993fg,Caprini:2015zlo}. 
When calculating the contribution of bubble collisions to the GW signal, we assume that, in the limit of large $\alpha$, the fraction of the latent heat deposited into the
bubble front is $\kappa_{\rm bc}\sim1$, leading to
\cite{Kosowsky:1991ua,Huber:2008hg,Caprini:2015zlo}
\bea
\label{Eq: bubble collision}
    h^2\Omega_{\rm bc}(f)&\approx&  \frac{(4.88 \!\times\! 10^{-6}) \left({f}/{f_{\rm bc}}\right)^{2.8}}{1+2.8\,\left({f}/{f_{\rm bc}}\right)^{3.8}}\!\left(\!\frac{H_{\rm RH}}{\beta}\!\right)^{\!2}\!\! \left(\frac{100}{g_*}\right)^{\!\frac13}  \ \ \ \ \ \ 
\eea
with  the peak frequency $f_{\rm bc}$,
\bea
\label{f1}
f_{\rm bc} &\approx& (3.7\times10^{-5} \ {\rm Hz})\left(\frac{g_*}{100}\right)^{\!\frac16}\!\left(\frac{\beta}{H_{\rm RH}}\right) \!\left(\frac{T_{\rm RH}}{1 \ {\rm TeV}}\right) .  
\eea
When considering cases in which the released energy is efficiently transferred to the plasma in the form of sound waves, we assume $\kappa_{\rm sw}\sim1$, and  %\cite{Espinosa:2010hh}, and
the resulting spectrum is \cite{Hindmarsh:2013xza,Caprini:2015zlo}
\bea\label{sw_contr}
h^2 \Omega_{\rm sw}(f) \approx  \frac{(1.86 \times 10^{-5})\left({f}/{f_{\rm sw}}\right)^3}{\left[1+0.75\,\left({f}/{f_{\rm sw}}\right)^2\right]^{7/2}}\!\left(\!\frac{H_{\rm RH}}{\beta}\!\right) \left(\frac{100}{g_*}\right)^{\frac13}\!\!\! , \ \ \ \ \ \ \ 
\eea
where 
the peak frequency, $f_{\rm sw}$,  is 
\bea
\label{f2}
f_{\rm sw} &\approx& (1.9\times10^{-4} \ {\rm Hz})\left(\frac{g_*}{100}\right)^{\frac16}\left(\frac{\beta}{H_{\rm RH}}\right) \left(\frac{T_{\rm RH}}{1 \ {\rm TeV}}\right)  \, . \ \ \  \ 
\eea
Implicit in this spectrum is an infinite sound wave lifetime,
$\tau_{\text{sw}}$. Note that this is a good approximation only if  turbulence and other damping processes are ignored, which is what we assume in this study.\footnote{The finite sound wave lifetime, $\tau_{\text{sw}}$,
results in a multiplicative factor that is a function of $\tau_{\text{sw}}$ and the expansion rate of the Universe when the sound waves were active~\cite{Ellis:2020awk,Guo:2020grp}. Currently, the value of $\tau_{\text{sw}}$ remains highly uncertain, though an analytical estimate is usually
adopted in the literature. Since this effect is simply an extra overall
factor, it could be taken as an additional parameter in the Bayesian inference.} Furthermore, in the supercooling limit, the $\alpha$ dependence vanishes from the spectra.\footnote{The GW spectra in \cite{Kosowsky:1991ua,Kamionkowski:1993fg,Huber:2008hg,Hindmarsh:2013xza,Caprini:2015zlo} are simulated for FOPTs  not exhibiting large supercooling. In our analysis we assume that those results can be extrapolated to the region of large $\alpha$.} 

In the following we consider the cases of bubble collision or of sound waves separately, assuming that one of the two mechanisms dominates the GW generation during the FOPT.

\section{Constraints on supercooled phase transitions using LVK data}
\label{Sec:General constraints}
To place constraints on model parameters using public data from LVK's first three observing runs \cite{PhysRevD.104.022004}, we apply a Bayesian search following the methodology of \cite{Romero_2021}. Although a search for FOPT signals was already performed in \cite{Romero_2021}, we now apply this search under the assumption that the signal comes from a supercooled phase transition. This allows for a simplification of the GW spectra, and therefore, less parameters in the Bayesian inference search.
%, resulting in an increased ease of reusability of the obtained upper limits.

The likelihood reads
\begin{equation} \label{eq:PE_likelihood}
p(\hat{C}^{IJ}(f)|\mathbf{\theta}) \propto \exp \!\!\left[ -\frac{1}{2} \sum_{IJ} \sum_f \left( \frac{\hat{C}^{IJ}(f) - \Omega_{\rm GW}(f|\mathbf{\theta})}{\sigma^2_{IJ}(f)}\right) \!\right]\!,
\end{equation}
where the sum runs over the detector baselines  $IJ$ and the frequencies $f$. The discrete set of frequencies is obtained by splitting the timeseries data into segments, and optimally combining them using inverse noise weighting. The cross-correlation estimator $\hat{C}^{IJ}(f)$ for the GW background using data from detectors $I$ and $J$ and the variance $\sigma^2_{IJ}(f)$ are data products of the LVK isotropic stochastic analysis \footnote{The data products are available to the public \cite{O3data}.} \cite{romanocornish,PhysRevD.104.022004}.

Below, two approaches will be explored. One based on approximating the GW background signal as a broken power-law and another one using the spectra provided in the previous section (see Eqs.\,\eqref{Eq: bubble collision} and \eqref{sw_contr}). In both cases, the contribution from the astrophysical background, i.e., from unresolved compact binary coalescences (CBCs), will be taken into account as well. For the frequency range we consider, the CBC background is expected to follow a power-law
\begin{equation}
    \Omega_{\rm cbc}(f)=\Omega_{\rm ref}\left(\frac{f}{f_{\rm ref}}\right)^{2/3},
\end{equation}
where $f_{\rm ref}$ is a reference frequency set to $f_{\rm ref}=25$ Hz \cite{PhysRevD.104.022004}.
It is worth noting that the constraints obtained in this section are general and  can be applied to any model exhibiting supercooling to constrain the underlying physical parameters. This will be done in Sec.~\ref{Sec:Models} for two concrete particle physics models.

\begin{figure}[t!]
\includegraphics[scale=.45]{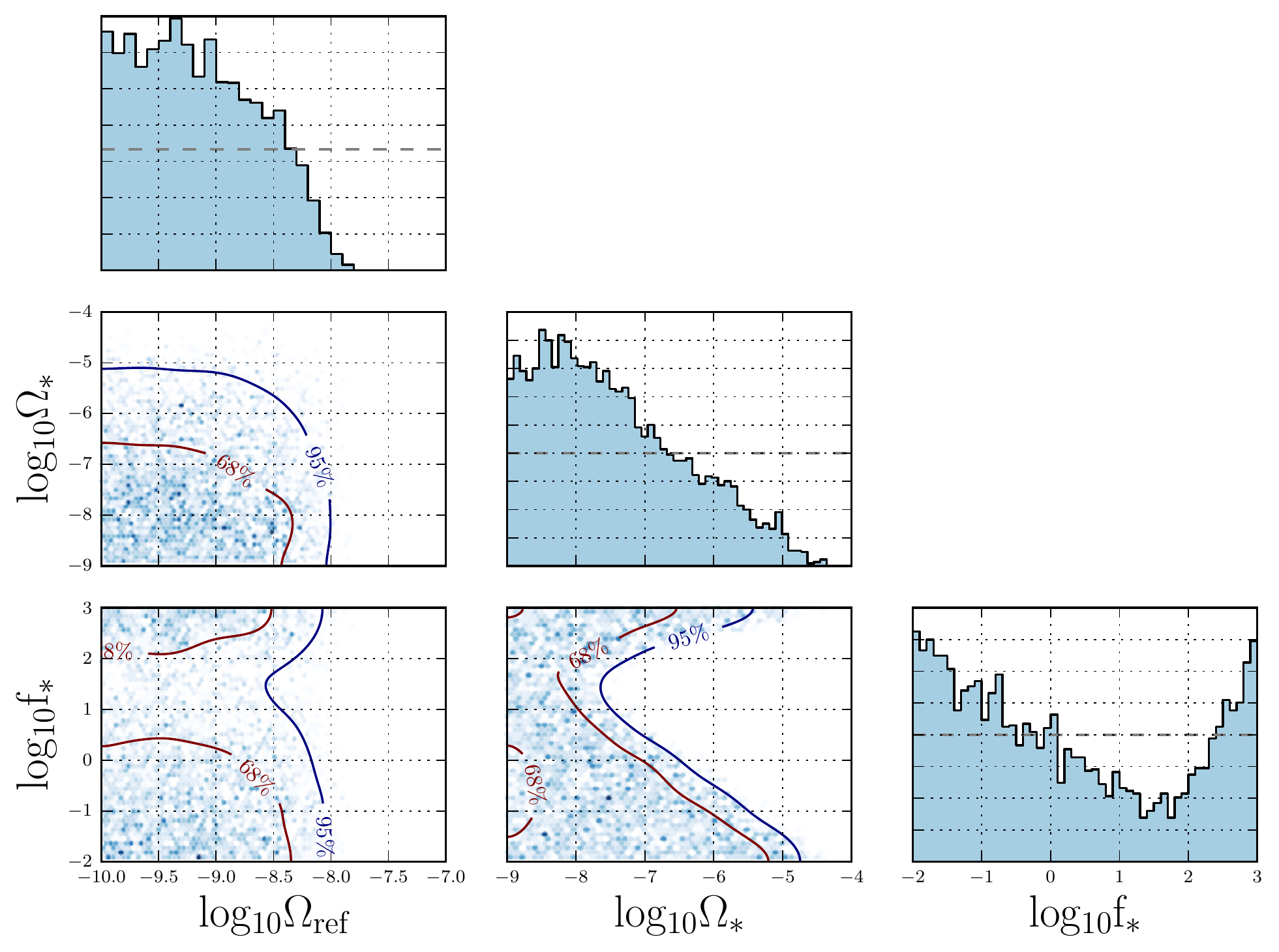}
\includegraphics[scale=.45]{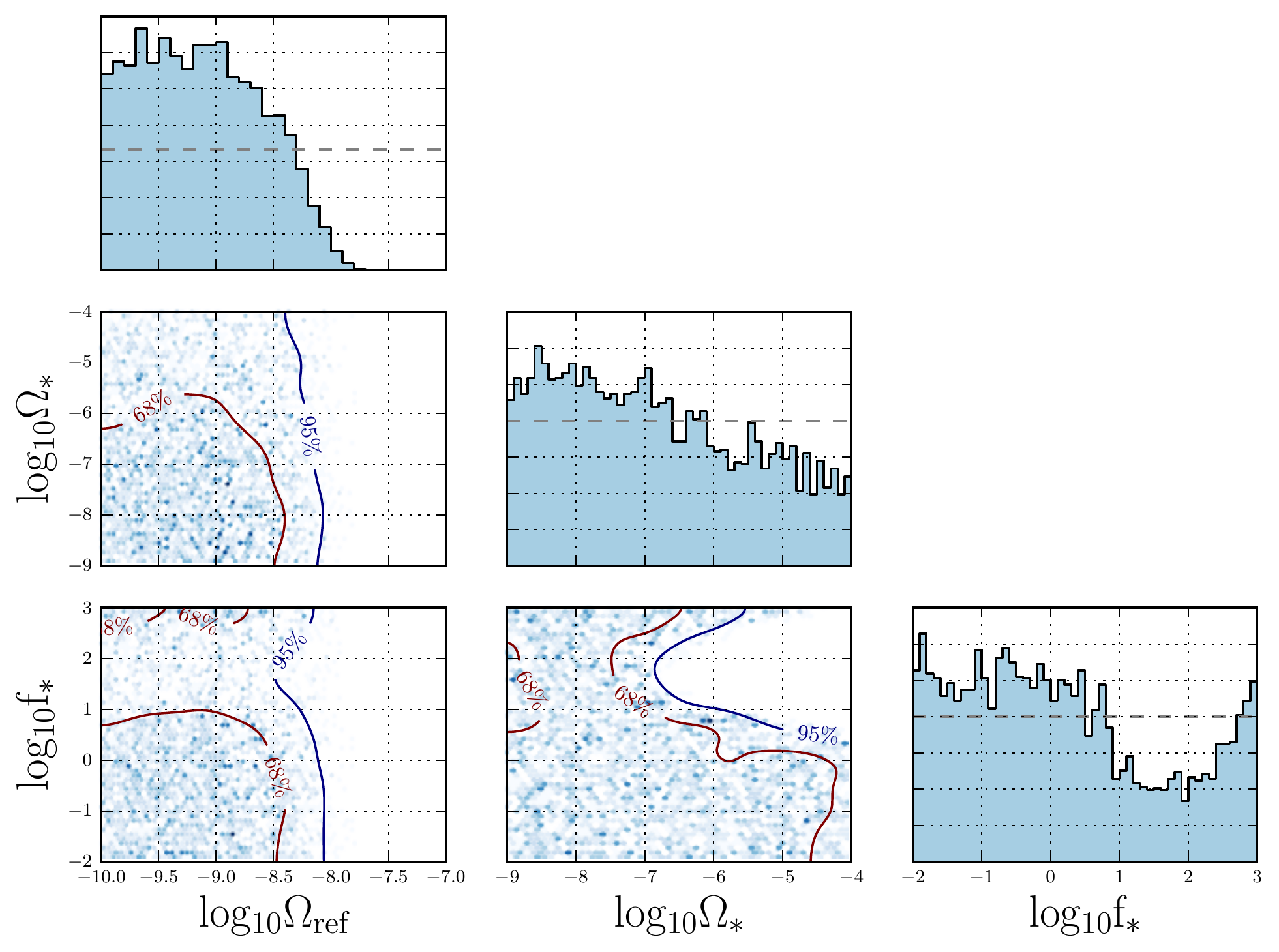}
\caption{Constraints from LVK O3 data on the broken power-law parameters of a FOPT signal, together with the contribution from the CBC background, assuming dominant bubble collision spectrum (top) and a dominant sound waves spectrum (bottom).}
\label{Fig:BPL LVK search}
\end{figure}

\subsection{General broken power-law search}

To constrain a GW background from FOPTs, we model the FOPT contribution to the GW spectrum with a broken power-law as
\begin{equation}
    \label{eq:BPL}
    \Omega_{\rm bpl}(f) =
              \Omega_*~\Big(\frac{f}{f_*}\Big)^{n_1}~ \Bigg[1+\Big(\frac{f}{f_*}\Big)^{\Delta}\Bigg]^{(n_2-n_1)/\Delta},
\end{equation}
where $n_1$ and $n_2$, respectively, denote the spectral indices before and after the peak, $\Delta$ is a peak smoothing parameter, and $\Omega_*$ and $f_*$ can be related to the peak amplitude and peak frequency of the spectrum. Note that the spectra introduced in Eqs.~\eqref{Eq: bubble collision},  \eqref{sw_contr}, approximately follow a broken power-law with parameters $n_1=3$, $n_2=-1$, $\Delta=4$ and $n_1=3$, $n_2=-4$, $\Delta=2$ for bubble collisions and sound waves, respectively. We perform a parameter estimation search for both contributions separately, corresponding to the case where one of them dominates the GW spectrum. In each case, the values of $n_1$, $n_2$, and $\Delta$ are set to the relevant values of that contribution, as given above. Note that in \cite{Romero_2021}, $n_2$ was allowed to vary.

\begin{table}[t!]
    \centering
    \begin{tabular}{ |p{.8cm}||p{3cm}|p{.97cm}||p{3cm}| }
    \hline
 \multicolumn{2}{|c|}{Broken power-law}&\multicolumn{2}{|c|}{Phenomenological} \\
 \hline
 $\Omega_{\rm ref}$&LogU[$10^{-10}$, $10^{-7}$]&$\Omega_{\rm ref}$&LogU[$10^{-10}$, $10^{-7}$]\\
 $\Omega_*$&LogU[$10^{-9}$, $10^{-4}$]&$\beta/H_{\rm RH}$&LogU[1,$10^3$]\\
 $f_*$&LogU[$10^{-2}$, $10^3$]&$T_{\rm RH}$&LogU[$10^5$,$10^{10}$]\\
 \hline
\end{tabular}
    \caption{Summary of the priors used for parameter estimation for the broken power-law model search and the phenomenological model search, where LogU stands for a log-uniform prior. The narrow prior on $\rm \Omega_{ref}$ stems from estimates of the CBC background~\cite{PhysRevLett.120.091101}. The peak frequency is chosen such that it lies in the region of highest sensitivity in LIGO-Virgo. Values lower than 1 for $\rm \beta/H_{\rm RH}$ are not considered, since otherwise the phase transition would not take place.
    \vspace{-4mm}}
    \label{tab:priors}
\end{table}

The likelihood to perform this search is given by Eq.\,\eqref{eq:PE_likelihood}, where $\Omega_{\rm GW}=\Omega_{\rm cbc}+\Omega_{\rm bpl}$. The GW parameters to be constrained are $\mathbf{\theta}_{\rm GW}=(\Omega_{\rm ref},~\Omega_*,~f_*)$ with priors given in Table~\ref{tab:priors} and results shown in Fig.~\ref{Fig:BPL LVK search}  for bubble collisions (top panel) and  sound waves (bottom panel). From the posteriors of the amplitude of the CBC background, $\rm \Omega_{ref}$, upper limits (ULs) at 95\% confidence level (CL) are obtained. The value for the case in which bubble collisions dominate is $5.60\times 10^{-9}$, which is consistent with the upper limit obtained in \cite{PhysRevD.104.022004,Romero_2021}. The UL in the case when sound waves dominate is also consistent with previous searches, with a value $5.70\times 10^{-9}$. Similarly, 95\% confidence level contours are obtained on the amplitude and peak frequency of the contribution from FOPTs, $\rm \Omega_*$ and $f_*$, as depicted in Fig.~\ref{Fig:BPL LVK search}. The values of the Bayes factor are $\log\mathcal{B}^{\rm CBC+BC}_{\rm noise} = -1.26$ and $\log\mathcal{B}^{\rm CBC+SW}_{\rm noise} = -0.80$, showing no evidence for a FOPT signal in the data.

\subsection{Phenomenological search}
\begin{figure}[t!]
\includegraphics[scale=.45]{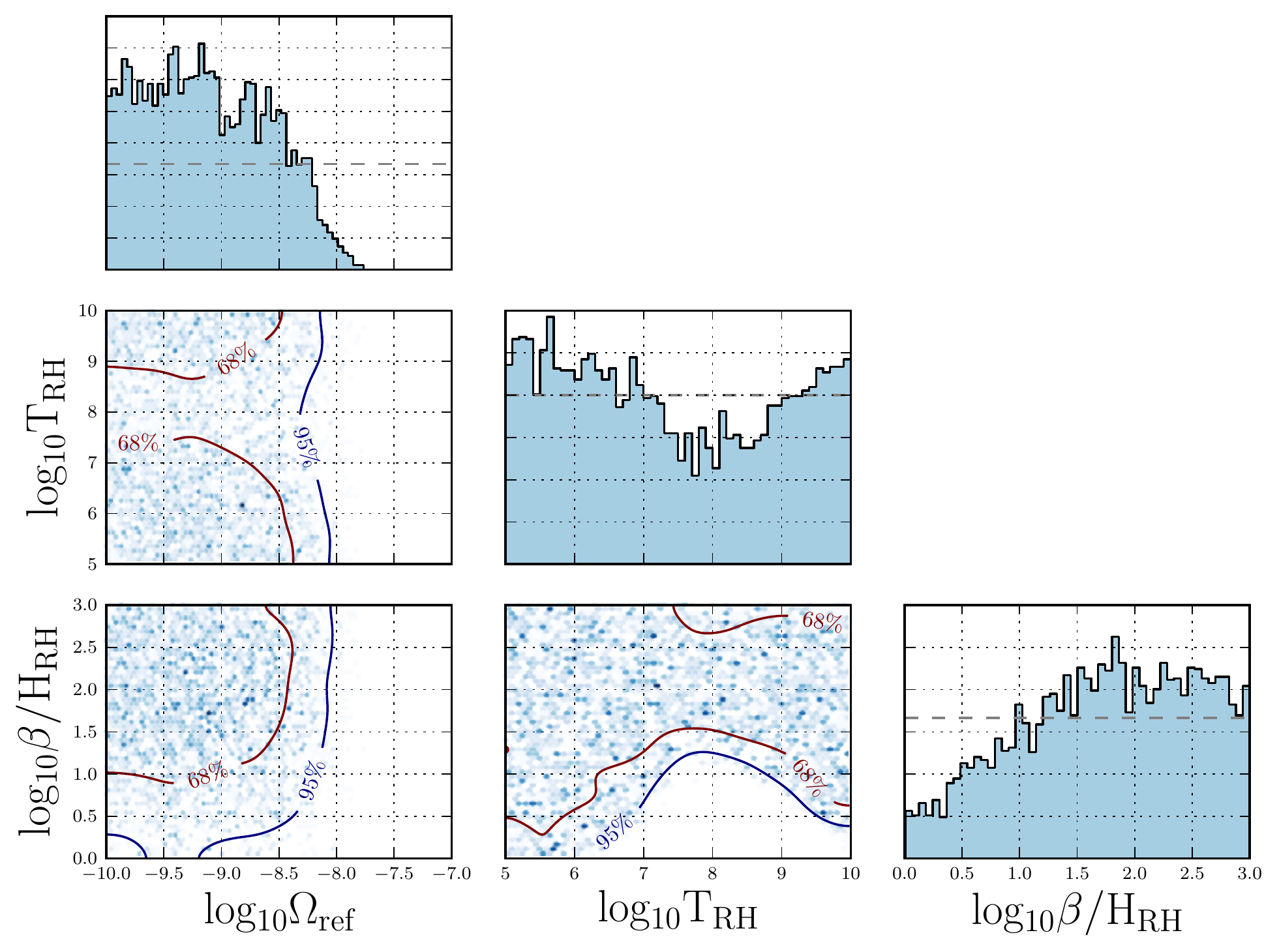}
\includegraphics[scale=.45]{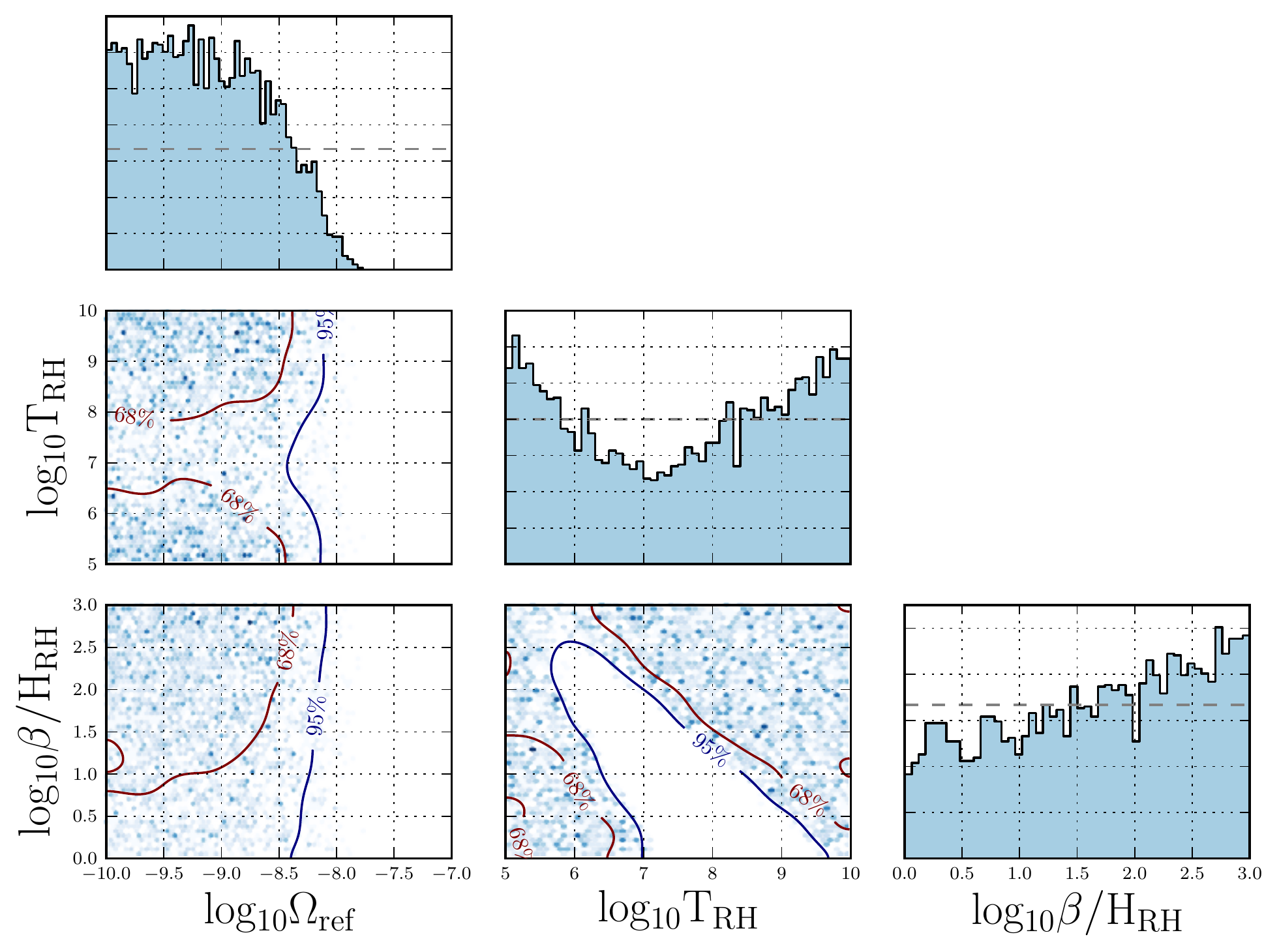}
\caption{Constraints from LVK O3 data on the phenomenological parameters $\beta/H_{\rm RH}$ and $T_{\rm RH}$ of a supercooled FOPT signal, together with the contribution from the CBC background, assuming a dominant bubble collision spectrum (top) and a dominant sound waves spectrum (bottom).}
\label{fig:pheno_o3}
\end{figure}

We now proceed with a different model assumption. Instead of the general broken power-law model used above, we consider the GW spectra introduced in Sec.\,\ref{Sec:GW}, more specifically Eqs.~\eqref{Eq: bubble collision} and \eqref{sw_contr}, corresponding to bubble collisions and sound waves, respectively. The likelihood used to perform this search is given by Eq.~\eqref{eq:PE_likelihood}, with $\Omega_{\rm GW}=\Omega_{\rm cbc}+\Omega_{\rm bc}$ and $\Omega_{\rm GW}=\Omega_{\rm cbc}+\Omega_{\rm sw}$ for bubble collisions and sounds waves, respectively. Therefore, the GW parameters to be constrained in this search are $\mathbf{\theta}_{\rm GW}=(\Omega_{\rm ref},~\beta/H_{\rm RH},~T_{\rm RH})$. We again highlight the difference with the search conducted in \cite{Romero_2021}, where the $\alpha$ parameter was included. As discussed earlier, for supercooled FOPTs, for which $\alpha\gg1$, neglecting this parameter is a valid assumption. The priors on the parameters used for parameter estimation are given in Table~\ref{tab:priors}, and the resulting posterior distributions are presented in Fig.~\ref{fig:pheno_o3}. From the posteriors of the amplitude of the CBC background, $\rm \Omega_{ref}$, ULs at 95\% CL are obtained. The value for the case in which bubble collisions or sound waves dominate is $5.89\times 10^{-9}$ and $5.93\times 10^{-9}$, respectively. They are consistent with the upper limit obtained in \cite{PhysRevD.104.022004,Romero_2021}. Furthermore, exclusions at 95\% CL for temperatures $T_{\rm RH}$ and inverse duration of the FOPT $\beta/H_{\rm RH}$ are depicted in Fig.~\ref{fig:pheno_o3}.

Let us emphasise that the constraints derived above can be used in any model exhibiting supercooling. More precisely, once a model and its parameters are specified, one can compute the expected FOPT parameters $\beta/H_{\rm RH}$ and $T_{\rm RH}$ (or equivalently $\Omega_*$ and $f_*$) and compare them with the 95\% confidence UL provided here. In this way, one uses GW data to exclude regions of the parameter space in concrete particle physics models. We will illustrate this in the next section for two  particle physics models.

\section{Two well-motivated particle physics models}
\label{Sec:Models}

The phenomenon of supercooling occurs in theories with
Coleman-Weinberg-type symmetry breaking \cite{PhysRevD.7.1888} or
strong coupling.  Several models of this type have been investigated in the literature in light of their enhanced GW signals \cite{Ellis:2019oqb,DelleRose:2019pgi,Marzo:2018nov,Jinno:2016knw,Lewicki:2020azd,Agashe:2019lhy,VonHarling:2019rgb,Ellis:2020awk,Lewicki:2021xku,Prokopec:2018tnq,Jinno:2019jhi,Baratella:2018pxi,Ellis:2020nnr}. In this study, we focus on  Model I \cite{Ellis:2019oqb} and Model II \cite{DelleRose:2019pgi}, which exhibit approximate conformal symmetry. They are both well-motivated from a particle physics point of view and have a minimal particle content. We note, however, that our analysis can be applied to any other model  with  supercooling. The general goal is to assess the detectability of signals from supercooled FOPTs with the LVK detectors, and determine the regions of parameter space that can be excluded with current GW data.

\subsection{Model I}

\subsubsection{Theoretical framework} 
The first model we consider is based on a theoretically attractive minimal ${\rm U}(1)_{B-L}$ extension of the SM gauge group \cite{Marzo:2018nov,Ellis:2019oqb,Ellis:2020nnr}.
  Upon introducing three right-handed neutrinos, the theory is anomaly-free, realises the seesaw mechanism,  and can be incorporated into  ${\rm SO}(10)$ grand unification.  
  The model includes only two new bosonic fields: a real scalar $\phi$ and a gauge boson $Z'$.
  
The zero temperature  scalar potential is given by
\bea
&&\hspace{-5mm}V_{0}(\phi) = \frac14{\lambda_\phi}\phi^4\nonumber\\ &+& \sum_{i=\phi,G,Z'} \frac{n_i}{64\pi^2}\left\{m_i^4(\phi)\left[\log\left(\frac{m_i^2(\phi)}{\mu^2}\right)-c_i\right]\right\} \ , \ \ \ \ \ 
\eea
where  $n_i$ is the number of degrees of freedom, $c_{\phi,G}=3/2$, $c_{Z'}=5/6$, $\mu$ is the renormalization scale, and $G$ denotes the Goldstone boson. The  field-dependent   masses  are:
\bea
&&m_{Z'}^2(\phi) = 4g^2\phi^2 \ , \  \ \ \ m_\phi^2(\phi) = 3\lambda_\phi \phi^2 \ , \nonumber\\
&&m_G^2(\phi) = \lambda_\phi \phi^2 \ ,
\eea
where $g$ is the gauge coupling.
The finite temperature part of the effective potential is
\bea\label{finiteT}
&&\hspace{-5mm}V_{\rm T}(\phi, T) \nonumber\\ 
&=& \ \frac{T^4}{2\pi^2} \sum_{i=\phi,G,Z'}\! n_i \!\int_0^\infty \!\!dy \,y^2 \log\left(1- e^{-\sqrt{m_i^2(\phi)/T^2 + y^2}}\right)\nonumber\\
&+& \ \frac{T}{12\pi} \sum_{j=\phi,G,Z_L'} \!n'_j \,\Big\{m_j^3(\phi) - [m_j^2(\phi) + \Pi_j(T)]^{\frac32}\Big\} ,
\eea
with the  thermal masses  given by
\begin{eqnarray}
 &&\Pi_\phi(T)=\Pi_G(T)= \left(g^2+\tfrac13\lambda_\phi\right)T^2 \ ,\nonumber\\
&& \Pi_{Z'_L}(T)  = 4 g^2T^2 \ ,
\end{eqnarray}
where the subscript $L$ denotes 
longitudinal components. 
\vspace{1mm}

This model has only  two free parameters relevant for the GW signal: the vacuum expectation value $v$ of the scalar field $\phi$, and the ${\rm U}(1)_{B-L}$ gauge coupling $g$. Trading $v$ for the the gauge boson mass $m_{Z'}$, related via
\bea
m_{Z'}={2g v} \ ,
\eea 
the two parameters describing Model I are $(m_{Z'},g)$. 

\subsubsection{Constraints from LVK O1+O2+O3 data} 
For each point $(m_{Z'},g)$ of the parameter space, one can compute the parameters describing the phase transition, i.e., $T_{\rm RH}$ and $\beta/H_{\rm RH}$, and the resulting GW spectrum. We  restrict ourselves to  $m_{Z'}\in[10^4,~10^{11}]$ GeV and $g\in[0.3,~0.4]$, which corresponds to FOPTs where the GW signal is dominated by sound waves and, therefore, given by Eq.~\eqref{sw_contr} \cite{Ellis:2020nnr}. If the gauge coupling $g$ is chosen to be larger than 0.4, the FOPT is not supercooled and $\alpha\sim1$.
Furthermore, we are not exploring  values of $g$ below 0.3, as these correspond to a regime where both bubble collisions and sound waves contribute considerably to the GW spectrum, as discussed in \cite{Ellis:2020nnr}.

\begin{figure}[t!]
\includegraphics[scale=.45]{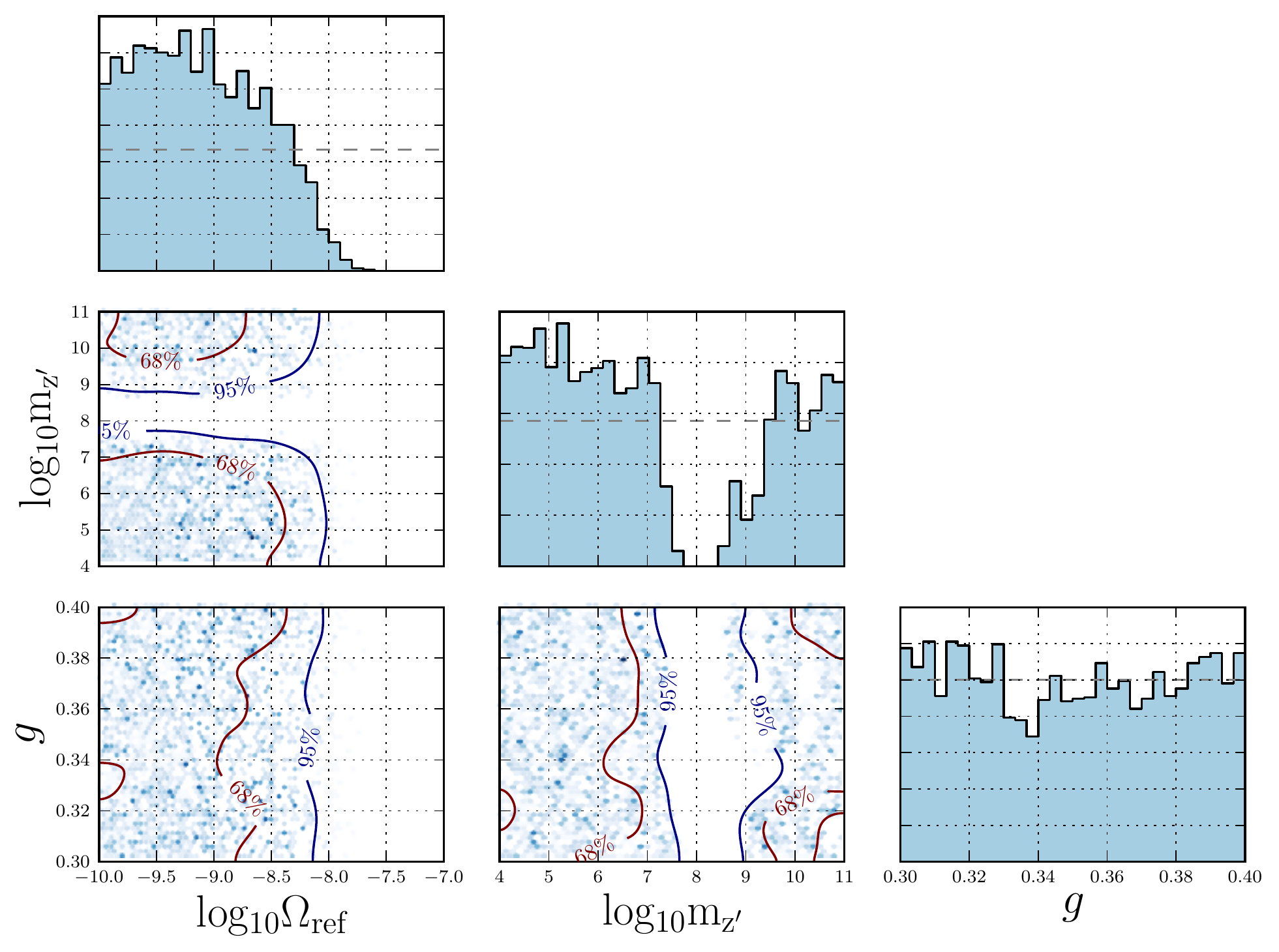}
\includegraphics[scale=.45]{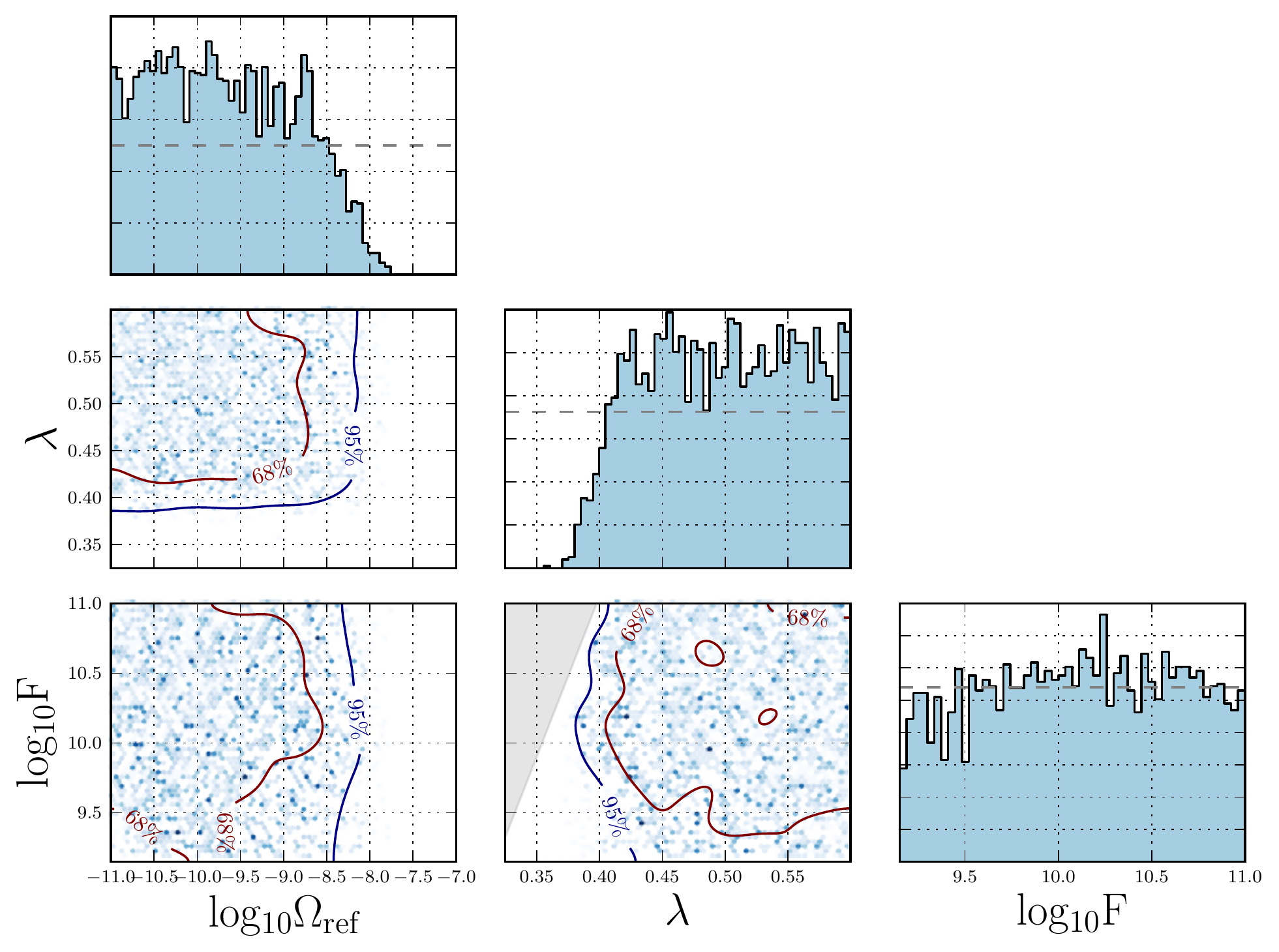}

\caption{Constraints on the parameter space $(m_{Z'},g)$ for Model I (top panel) and on the parameters $\lambda$ and $F$ for Model II (bottom panel), together with constraints on the astrophysical CBC background amplitude $\Omega_{\rm ref}$ using LVK O3 data. The gray region in the bottom plot corresponds to a region where nucleation does not occur and the phase transition does not complete.}
\label{Model12}
\end{figure}

\begin{table}[t!]
    \centering
    \begin{tabular}{ |p{.8cm}||p{3.1cm}|p{.8cm}||p{3.8cm}| }
    \hline
 \multicolumn{2}{|c|}{Model I}&\multicolumn{2}{|c|}{Model II} \\
 \hline
 $\Omega_{\rm ref}$&LogU[$10^{-10}$, $10^{-7}$]&$\Omega_{\rm ref}$&LogU[$10^{-10}$, $10^{-7}$]\\
 $m_{Z'}$&LogU[$10^{4}$, $10^{11}$] (GeV)&$F$&LogU[$1.4$$\times$$10^9$,\,$10^{11}$] (GeV)\\
 $g$&U[0.3, 0.4]&$\lambda$&U[$0.325$, $0.6$]\\
 \hline
\end{tabular}
    \caption{Summary of the priors used for parameter estimation for Model I and Model II, where U stands for a uniform and LogU for a log-uniform prior. The narrow prior on $\rm \Omega_{ref}$ stems from estimates of the CBC background~\cite{PhysRevLett.120.091101}.}
    \label{tab:priorsModels}
\end{table}

We perform a parameter estimation search over this parameter space, and include the contribution of the CBC background. The likelihood is given by Eq.~\eqref{eq:PE_likelihood}, with $\Omega_{\rm GW}=\Omega_{\rm cbc}+\Omega_{\rm sw}$, where $\Omega_{\rm sw}$ is calculated from Eq.~\eqref{sw_contr} using the model parameters $(m_{Z'},g)$. Thus, the parameters of the search are $\mathbf{\theta}_{\rm GW}=(\Omega_{\rm ref},~m_{Z'},~g)$. The priors are summarised in Table \ref{tab:priorsModels} and the results are shown in  Fig.\,\ref{Model12} (upper panel) which depicts the resulting posteriors. The upper limits on the amplitude of the astrophysical CBC background are consistent with \cite{PhysRevD.104.022004,Romero_2021}. Furthermore, a region of parameter space  around $m_{Z'}\sim\mathcal{O}(10^8\text{ GeV})$ is excluded, and corresponds to FOPT GW signals peaked within the frequency range of the LVK detectors.

\begin{figure}[t!]
\includegraphics[scale=.48]{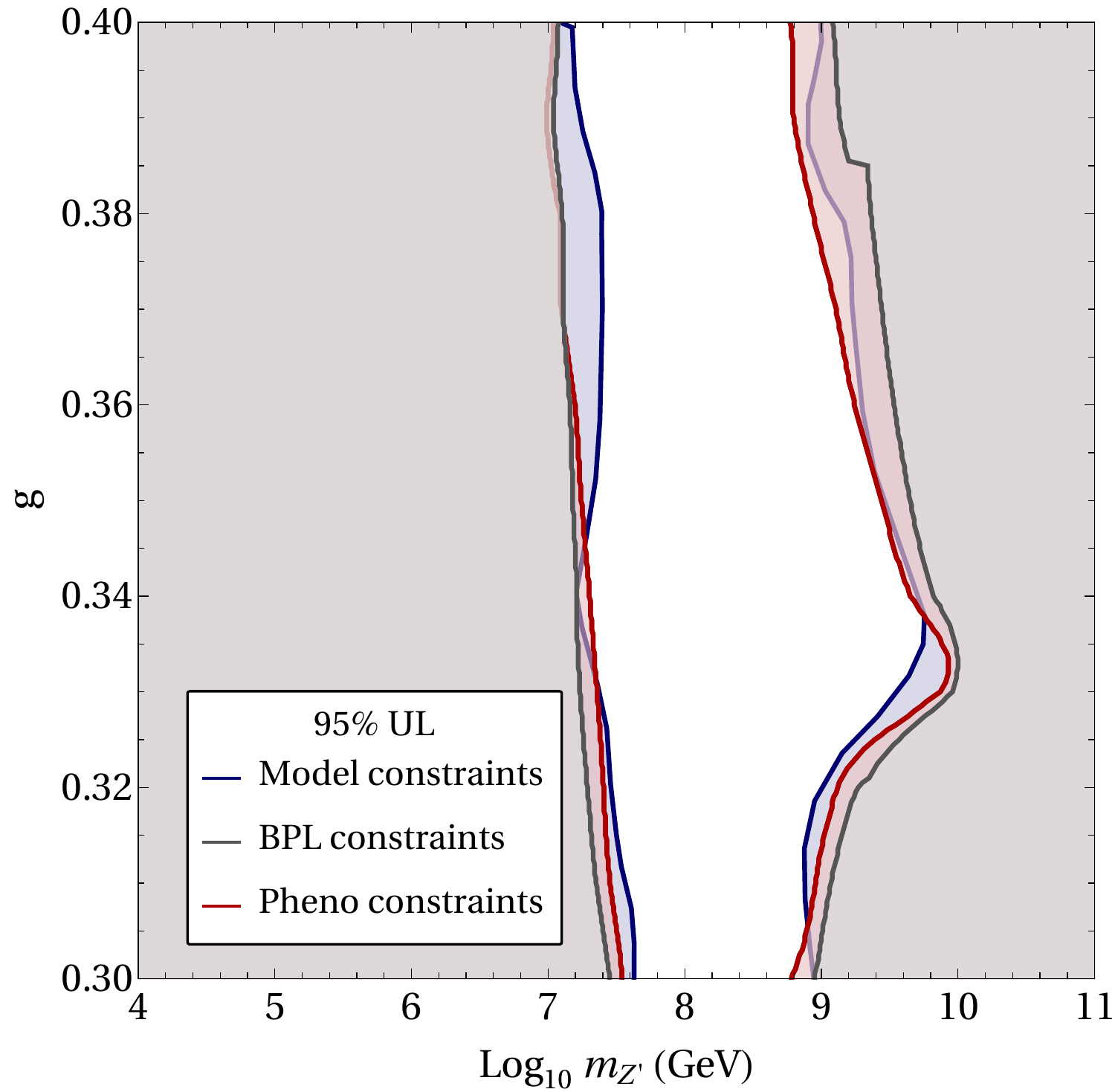}
\includegraphics[scale=.5]{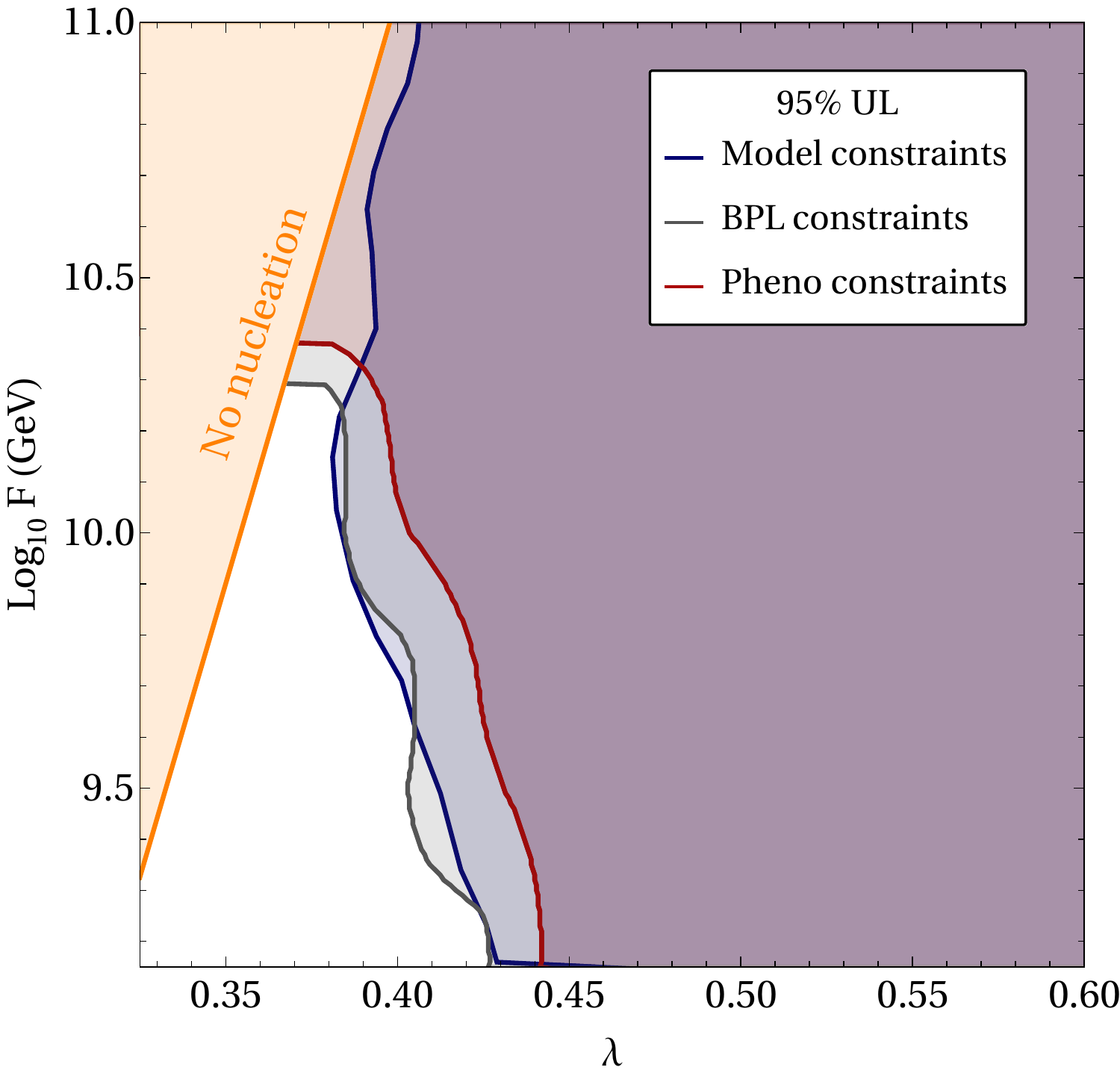}

\caption{Comparison of the constraints on the parameter space $(m_{Z'},g)$ for Model I (top panel) and on $(\lambda,F)$ for Model II (bottom panel) obtained by constraining the model parameters directly as in Fig. \ref{Model12} (blue line), with those obtained by adopting the BPL model as given in Fig. \ref{Fig:BPL LVK search} (gray line), and those adopting the phenomenological model in Fig. \ref{fig:pheno_o3} (red line).}\label{ModelComparingUL}
\end{figure}

We now compare the exclusion regions obtained directly on the parameters of the model with the ones deduced from the analysis in Sec.~\ref{Sec:General constraints}~\footnote{The particle physics masses and couplings are independent and uncorrelated. Both the broken power-law search and the phenomenological search parameters (and their priors) can be mapped using these fundamental parameters.}.  Given a  choice of the parameters $(m_{Z'},g)$, one can verify whether the corresponding values of  $(\beta/H_{\rm{RH}},T_{\rm{RH}})$ or $(f_*,\Omega_*)$ are excluded using the search analysis in Sec.~\ref{Sec:General constraints}. As  Fig.~\ref{ModelComparingUL} demonstrates, a good agreement is found between the various exclusion regions, regardless of the search performed. 
Therefore, the results obtained in Sec. \ref{Sec:General constraints} are easily reinterpreted in any specific model with supercooling. This is also supported by the analysis we perform below for another well-motivated  particle physics model.

\subsection{Model II}

\subsubsection{Theoretical framework} 
This  model is based on a radiatively broken  ${\rm U(1)}$ Peccei-Quinn symmetry 
 \cite{DelleRose:2019pgi}, introduced to solve the strong CP problem, and leading to the appearance of a dark matter candidate -- the axion. 
It extends the SM by including just two new  complex scalar fields, $S$ and $X$, which are SM singlets, and both carry Peccei-Quinn charges.

The tree-level scalar potential is 
\bea
V_{\rm tree} = \lambda_S |S|^4 + \lambda_X |X|^4 + \lambda_{SX} |S|^2 |X|^2 \ . 
\eea
It  exhibits a flat direction for $\lambda_{SX} = -2\sqrt{\lambda_S \lambda_X}$, which can be parameterised by 
\bea
(S,X) = (\sin\alpha,\cos\alpha) \frac{\sigma}{\sqrt2} \ , \hspace{2mm} \sin^2\!\alpha = \frac{\sqrt{\lambda_X}}{\sqrt{\lambda_S}+\sqrt{\lambda_X}} \ .\ \ \ \ \ \ 
\eea
The mass of the field along the  direction orthogonal to $\sigma$ is
\bea
m_\tau = (4\lambda_S\lambda_X)^{1/4}\sigma \ .
\eea
Assuming that the condition for the flat direction holds at the renormalisation scale $\Lambda$, and switching the parameter $\Lambda$ for the field value at the minimum of the potential $F$,
the zero temperature scalar potential is given by 
\bea\label{fdm}
V_{0}(\sigma)=\frac{2\lambda_S\lambda_X}{16\pi^2}\sigma^4 \left(\log\frac{\sigma}{F}-\frac14\right) \ .
\eea
At the minimum $\sigma$ has a loop-suppressed mass, whereas the phase of $X$ is massless up to QCD anomalies, and becomes the axion  with 
a decay constant  $F_a=F\cos\alpha$. 
The finite temperature part of the effective potential is given by a formula analogous to  Eq.\,(\ref{finiteT}), but with just a single term  involving $m_\tau$.
To prevent  the 
finite temperature effects  from moving the true vacuum away from the flat direction, we set
\bea
\lambda_X=\lambda_S\equiv \lambda \ ,
\eea
which is equivalent to imposing a $\mathbb{Z}_2$ symmetry at the level of the  Lagrangian. As a result, Model II is described by just  two parameters:  $(\lambda, F)$.

\subsubsection{Constraints from LVK O1+O2+O3 data}
Similarly as for Model I, one can compute the FOPT parameters $\beta/H_{\rm{RH}}$ and $T_{\rm{RH}}$, and determine the GW spectrum. The ranges of parameters we consider are: $F\in[1.4 \times 10^9,~10^{11}]$ GeV and $\lambda\in[0.325, 0.6]$. A value of $F$ smaller than $1.4 \times 10^9$ GeV (corresponding to an axion decay constant of $10^9$ GeV) is experimentally excluded \cite{DiLuzio:2020wdo}, whereas   values of $\lambda$ lower than 0.325 correspond to cases when the phase transition does not complete, i.e., no nucleation occurs. The upper limits on $F$ and $\lambda$ are not constrained and were set arbitrarily in Fig.\,\ref{ModelComparingUL}.

We again conduct a parameter estimation  directly on the parameters of the model.  In the case of Model 2 the dominant GW contribution  comes from bubble collisions \cite{DelleRose:2019pgi}.
In the likelihood given by Eq.~\eqref{eq:PE_likelihood}, $\Omega_{\rm GW}=\Omega_{\rm cbc}+\Omega_{\rm bc}$, where $\Omega_{\rm bc}$ is given by Eq.\,\eqref{Eq: bubble collision} and can be obtained from the underlying model parameters $(\lambda,F)$. The parameters used for the search are $\mathbf{\theta}_{\rm GW}=(\Omega_{\rm ref},~F,~\lambda)$ and the priors on $\Omega_{\rm ref}$, $F$ and $\lambda$  are summarised in Table \ref{tab:priorsModels}. The lower panel in Fig.\,\ref{Model12} displays the exclusion regions implied by the current LVK O3 data. The gray region represents part of the parameter space where no nucleation occurs and the phase transition does not complete. As shown in Fig.\,\ref{Model12}, part of the parameter space can be excluded at a 95\% confidence level. This mostly puts constraints on the values of $\lambda$, excluding smaller values, as these are the ones that give rise to the strongest GW signals. Furthermore, one notes consistency with the usual CBC upper limits found in this work, and in \cite{PhysRevD.104.022004,Romero_2021}.

One can now compare the exclusion regions obtained directly on the model parameters, with those derived following the analysis in Sec.~\ref{Sec:General constraints}. The results are shown in Fig.~\ref{ModelComparingUL}, where we note an agreement between the exclusion regions arising from the different searches, similar to the agreement obtained in the case of Model I. Once again, this illustrates how the exclusion regions in Sec.~\ref{Sec:General constraints} can be used to constrain any supercooled FOPT at a particle physics model level.

\section{Detectability of a gravitational-wave background}
\label{Sec:PI}
In this section we  briefly compare various ways to address the detectability of a GW background.
Instead of using the full Bayesian inference run, one often uses power-law integrated (PI) sensitivity curves, proposed in \cite{Thrane_2013} as a graphical way to address the detectability of a GW background with a power-law dependence within the frequency band of the detector. 
%
%However, a GW background coming from FOPTs with spectra given in Sec.\,\ref{Sec:GW} would display a broken power-law behaviour. In addition, the PI method does not take into account the presence of the CBC background. Nevertheless, the PI curves are still widely used within the FOPT community to make statements about the detectability of such a signal. We investigate below the applicability of these PI curves to FOPTs. 
%
However, a GW background coming from FOPTs with spectra given in Sec.\,\ref{Sec:GW} would display a broken power-law behaviour. In addition, the presence of the CBC background
should be taken into account properly when assessing the experimental sensitivity to cosmological GW backgrounds.
In order to quantify the impact of these aspects,
we investigate below the applicability of the PI curves method to FOPTs.
[We refer the reader to the Appendix for a review of the construction of PI curves, as outlined in \cite{Thrane_2013}.]

We now discuss the FOPT detectability with the LVK detectors. Using the broken power-law model described in Sec.~\ref{Sec:General constraints}, we generate a GW background signal dominated by bubble collisions for a range of $\Omega_*, f_{*}$. If a resulting GW background signal is larger than a PI curve constructed, we consider this a detection at the curve's $\rho_{\rm thr}$ level. In addition, the SNR of each generated GW background signal is calculated. We compare these results to a Bayesian analysis. Again, assuming a domination of bubble collisions, 800 simulated signals over a range of $\Omega_*, f_{*}$ are injected assuming a combined FOPT+CBC background model with a CBC background amplitude at the reference frequency of $\Omega_{\rm{ref}} = 5.9 \times 10^{-9}$ ~\cite{Romero_2021}. We analyse assuming a pure CBC background, a FOPT signal and a combined FOPT+CBC signal. We subtract the CBC model Bayes factor from the FOPT+CBC Bayes factor to see the preference for the latter model over the former. This procedure is repeated assuming a signal dominated by sound waves, and our results are plotted in Fig.~5.
%\ref{BC_SW_DetStats}. 

\begin{figure}[t!]
\vspace{-4mm}
\includegraphics[scale=.18]{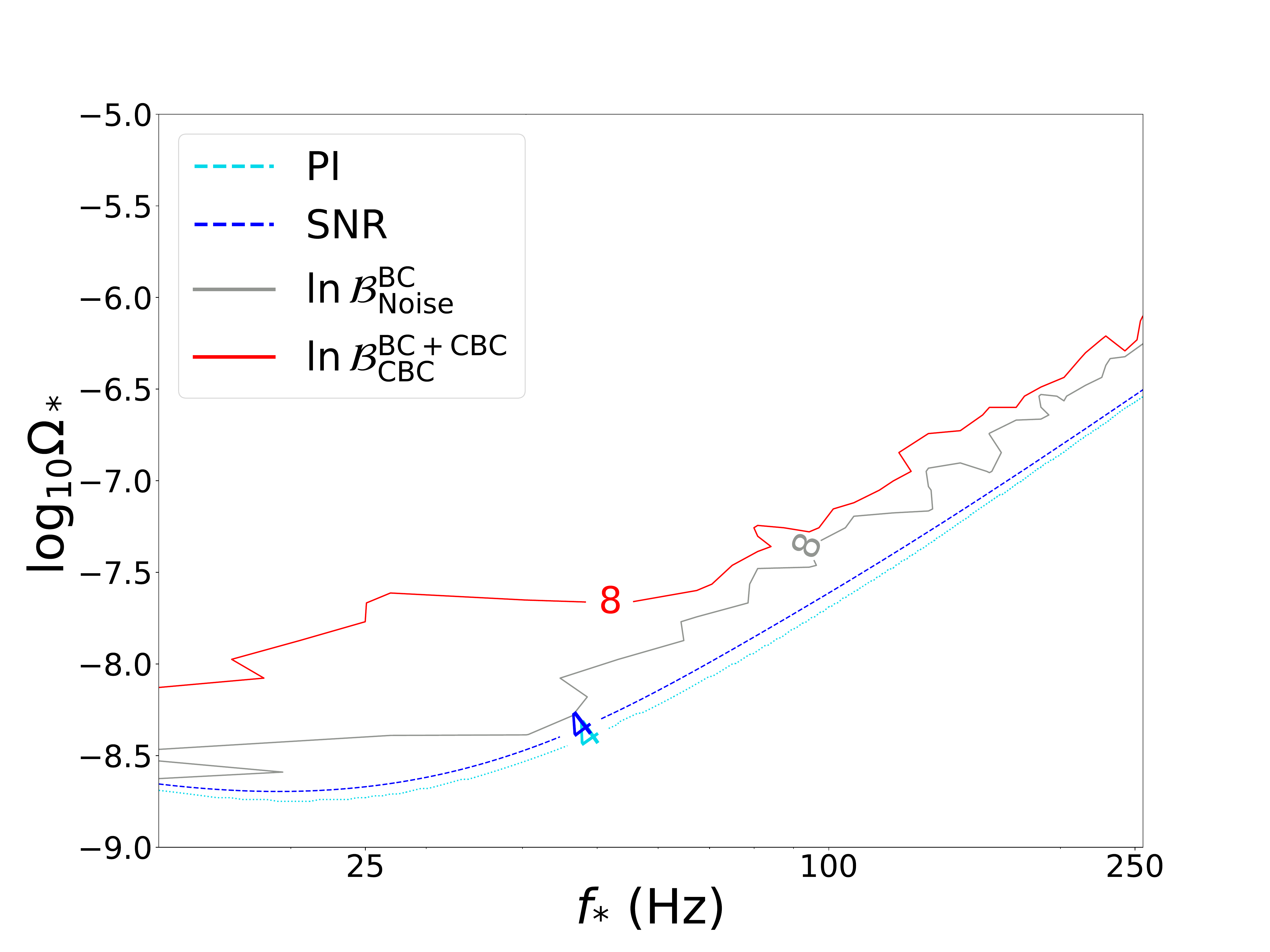}
\vspace{-5mm}
\includegraphics[scale=.18]{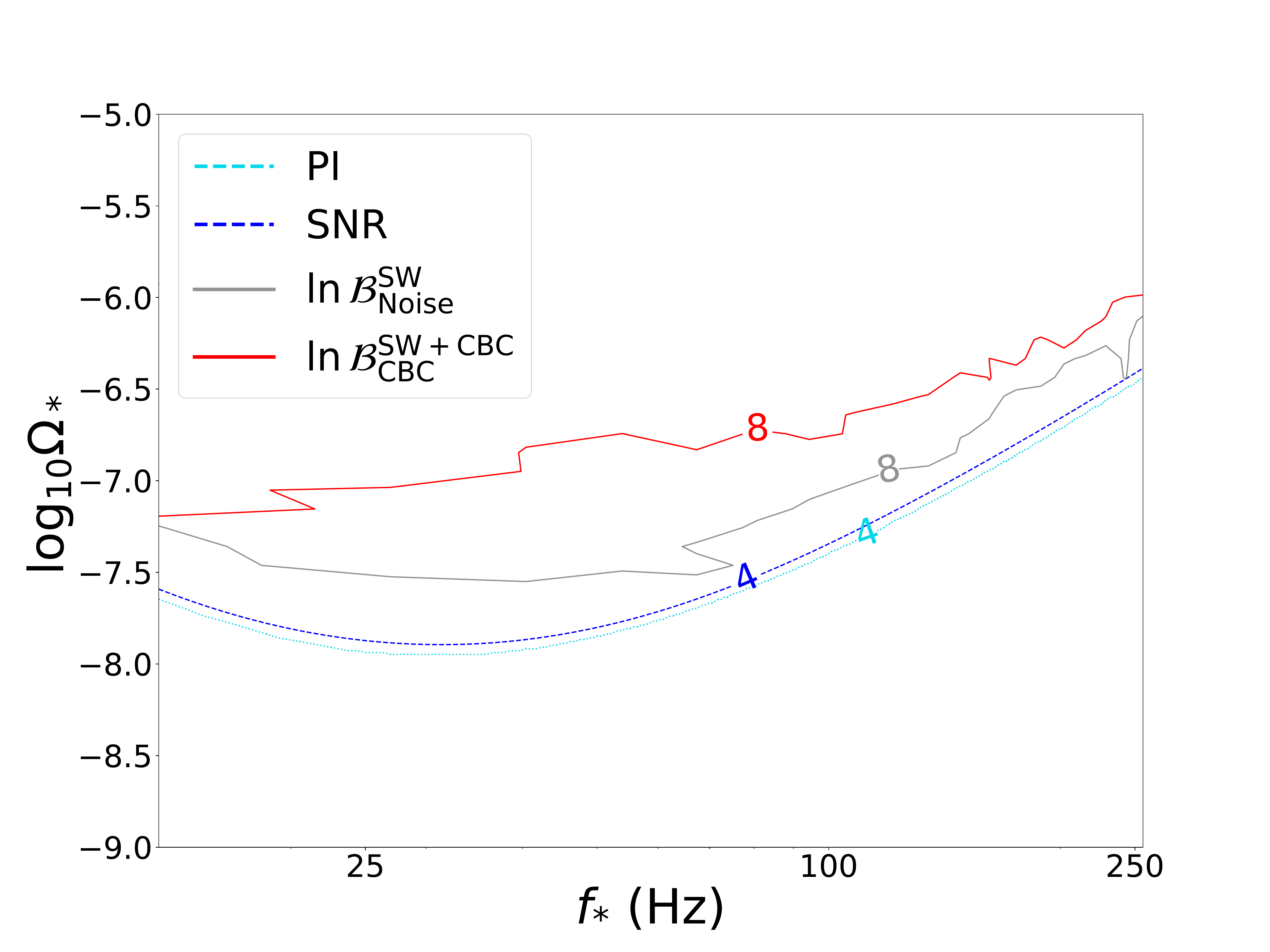}
\label{BC_SW_DetStats}
\vspace{-5mm}
\caption{Detectability region within the broken power-law parameter space using PI, SNR curves (dashed) and Bayesian runs (solid) assuming bubble collisions (top) and sound waves (bottom). We choose as detection threshold SNR = 4 and $\ln \mathcal{B}_{\rm{Noise}}^{\rm{FOPT}} =\ln \mathcal{B}_{\rm{CBC}}^{\rm{FOPT+CBC}} = 8$.}
\end{figure}

The calculated PI, SNR and Bayes factor curves follow similar trends. LVK detection capabilities improve for smaller $f_{*}$ and optimise at about 25 Hz, before increasing again. This is explained by the LVK network being most sensitive at this frequency, allowing a more optimistic outlook on broken power-law FOPT signals peaking at this frequency. The PI and SNR detection curves track each other very closely with the SNR curve being slightly more conservative. 
For values of $f_* \leq 250$ Hz,  using the PI method one can expect a detection at $\rho_{\rm thr} \geq 4$ for a bubble collision GW background with $\log_{10} \Omega_* \geq -6.54$ and $\log_{10} \Omega_* \geq -6.44$ for a sound wave GW background. Similarly, one can achieve $\rm{SNR}_{\rm{thr}} \geq 4$ for a bubble collision background with $\log_{10} \Omega_* \geq -6.50$, and $\log_{10} \Omega_* \geq -6.39$ for a sound wave background.
The SNR, PI curves are optimised with $\log_{10} \Omega_* \geq -8.70, -8.75$ for a bubble collision dominated background, and $\log_{10} \Omega_* \geq -7.89, -7.95$ for a sound wave dominated background when $f_{*}=25$ Hz.

Turning to the Bayesian analysis, we see that the resulting detection curves are more conservative than the PI and SNR ones. For a detection at Bayes factor $\ln \mathcal{B}_{\rm{Noise}}^{\rm{FOPT}} \geq 8$, a bubble collision GW background with $\log_{10} \Omega_* \geq -6.32$ and $\log_{10} \Omega_* \geq -6.23$ for a sound wave GW background is needed. To find a preference for a combined FOPT+CBC model over a CBC background model at Bayes factor $\ln \mathcal{B}_{\rm{CBC}}^{\rm{FOPT+CBC}} \geq 8$, a bubble collision GW background with $\log_{10} \Omega_* \geq -6.28$ and $\log_{10} \Omega_* \geq -5.98$ for a sound wave GW background is needed. 
Similar to the PI, SNR curves, the Bayes factor curves in both FOPT models are optimised when considering models with smaller $f_{*}$.
A stronger GW background signal is needed to find a preference for a combined FOPT+CBC model over a pure CBC background.

We conclude that a data-based Bayesian search for broken power-law signal, including the effect of the CBC background, has approximately one order of magnitude less sensitivity in $\Omega_*$ than the simple PI estimate, on the frequency range accessible to LVK.

\section{Future outlook}
\label{Sec:3g}

We complete our study by looking ahead and making projections for the sensitivity of 3G interferometers to a supercooled FOPT that could have occurred at energies inaccessible to particle colliders. The proposed Einstein Telescope (ET) \cite{Punturo:2010zz} and Cosmic Explorer (CE) \cite{Abbott_2017_3G,Reitze:2019dyk} are expected to extend our astrophysical horizon to distant redshift, revealing the majority of CBCs in the Universe. This will help subtract individual sources and reduce the astrophysical contribution to the GW background, in hope of revealing a cosmological background. 

We simulate 400 signals containing the residual CBC background \cite{Sachdev:2020bkk} and a bubble collision dominated supercooled phase transition for a range of $\beta/H_{\rm{RH}}$ and $T_{\rm{RH}}$ values. We then compute the log Bayes factor of a CBC+FOPT model to noise, and a CBC model only to noise; subtracting the two determines the preference for the presence of a FOPT signal in the data. The analysis is then repeated for the case of a dominant sound wave contribution to the FOPT signal. The 3G network used places ET at Virgo and two CEs at the Hanford and Livingston locations.

\begin{figure}[h!]
\includegraphics[scale=.4]{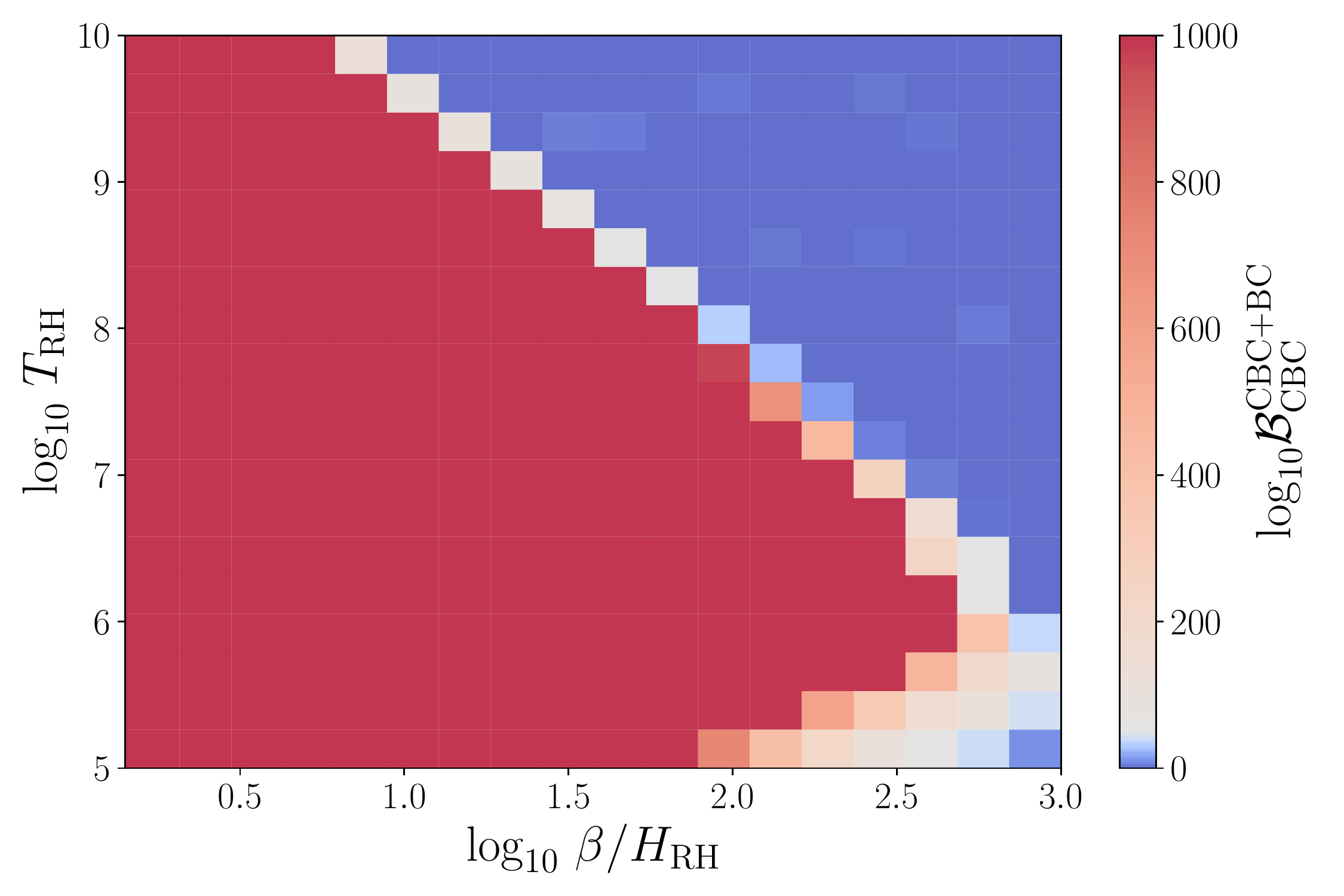}
\includegraphics[scale=.4]{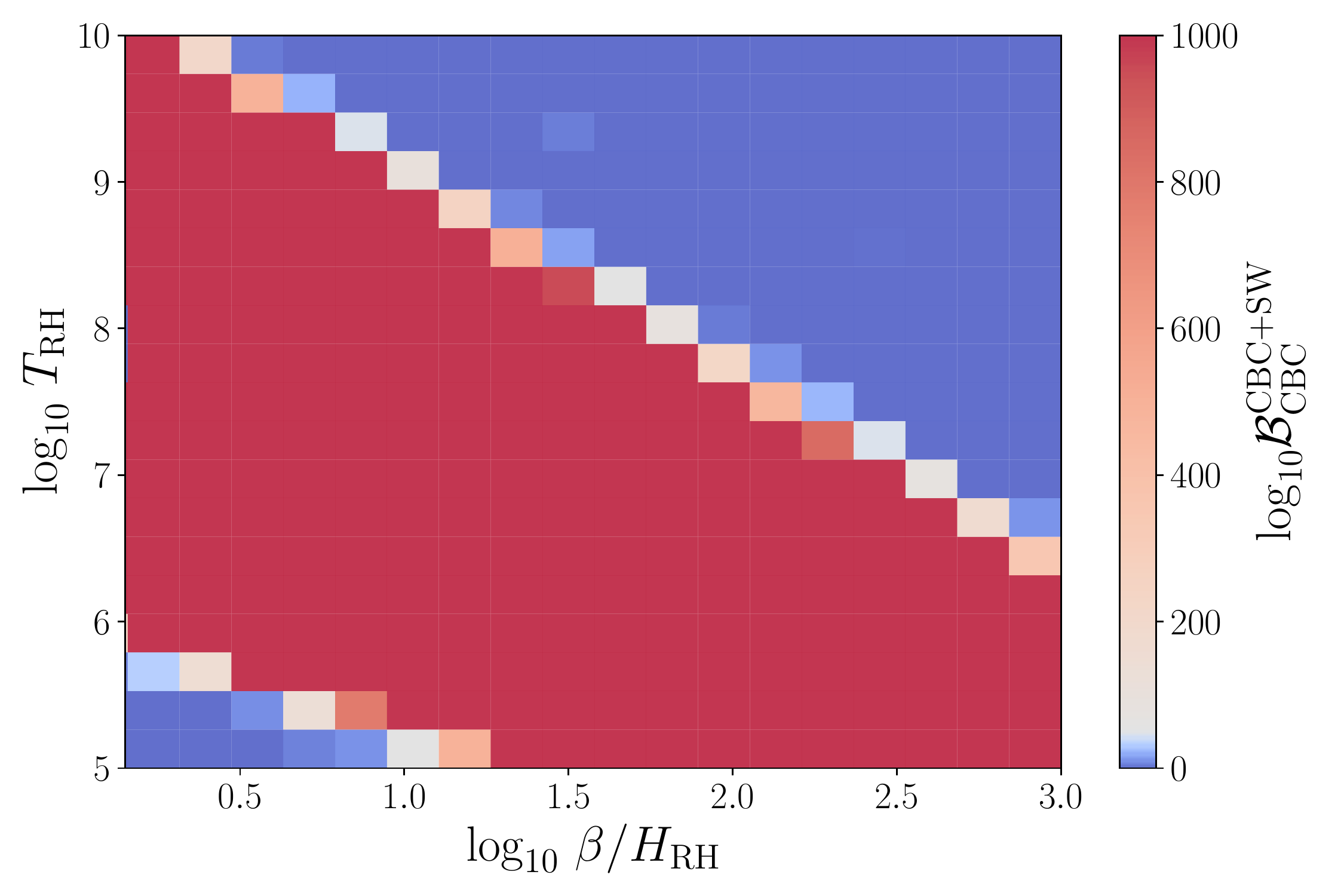}
\caption{Preference for a model containing a supercooled phase transition and an astrophysical CBC background over a model with an astrophysical background only. Injections of a bubble collision (BC) dominated FOPT show great constraining power of such a signal with a network of 3G detectors (top), and similarly for injections of a sound waves (SW) dominated FOPT (bottom).}
\label{fig:3g}
\end{figure}

Our results are presented in Fig.\,\ref{fig:3g}. 
With the future 3G detectors, we find that a significant part of the parameter space can be probed in both cases. Sound wave and bubble collision dominated supercooled phase transition scenarios are depicted on the top and bottom panels of Fig. \ref{fig:3g},  respectively.

\section{Conclusions}
Standard high energy physics experiments are  approaching limits of their discovery potential. In many cases, the most natural regions of model parameter space relevant for addressing questions in particle physics are not even within their target sensitivity. New discovery tools are needed to probe physics at the PeV energy scale and beyond. Such a novel and powerful  discovery tool has recently been provided by GW detectors, with their relevance destined only to increase in future years, given the upcoming upgrades to existing GW experiments and the construction of new detectors sensitive to a wider range of frequencies.

To demonstrate the huge opportunity for particle physics arising from GW searches, we carried out the pioneering study in which we used the data from the first three LVK  observing runs (O1, O2 and O3) to perform a Bayesian analysis and set direct limits on the parameter space of particle physics models. This is a natural extension of the previous work \cite{Romero_2021}, in which only general constraints on FOPT parameters were derived. In our analysis we focused on supercooled FOPTs, since they are naturally characterised by an enhanced signal strength, potentially already within the reach of current LVK detectors.

To show how the procedure works, we applied our analysis to two well-motivated particle physics models, which address some of the most intriguing open questions about the Universe: the dark matter puzzle, the strong CP problem, the origin of the neutrino masses, and unification of forces. We place the Bayesian 95\% upper limits on the parameter space of those models, providing valuable insight into the available room for new physics. The same  strategy can be used to impose limits on other models exhibiting supercooled FOPTs and is left for future work.

Apart from conducting the analysis using the available LVK O1-O3 data, we provide an outlook on the reach of 3G detectors. This methodology can also be applied to future LVK upgrades, as well as next generation detectors.
It is worth emphasizing  that our work bridges the gap between data analysis and phenomenological studies, making the constraints from GW searches easier to reinterpret, 
and applicable
to any particle physics model with  a supercooled phase transition.

\begin{acknowledgments}

This material is based upon work supported by NSF's LIGO Laboratory which is a major facility fully funded by the National Science Foundation.

The authors acknowledge computational resources provided by the LIGO Laboratory and supported by National Science Foundation Grants
PHY-0757058 and PHY-0823459. The software packages used in this study are \texttt{matplotlib}~\cite{Hunter:2007}, \texttt{numpy}~\cite{numpy},  \texttt{bilby}~\cite{Ashton:2018jfp}, \texttt{PyMultiNest}~\cite{refId0},  \texttt{HTCondor}~\cite{condor-practice}. The authors thank Michele Redi and Noam Levi for useful discussions, as well as  Tom Callister for the basis of the corner plot script.

K.M. is supported by King's College London through a Postgraduate International
Scholarship. B.F. is supported by the National Science Foundation under Grant No. PHY-2213144.
K.T. is supported by FWO-Vlaanderen through grant number 1179522N and A.S. through grant G006119N.  M.S. is supported in part by the Science and Technology Facility Council (STFC), United Kingdom, under the research grant ST/P000258/1. A.M., K.T. and A.S. are supported in part by the Strategic Research
Program ``High-Energy Physics'' of the VUB. H.G., F.Y. and Y.Z.
are supported by U.S. Department of Energy under Award No. DESC0009959.

The article has a  LIGO document number LIGO-P2200254 and an Einstein Telescope document number ET-0194A-22. %\bart{and a Virgo document number Virgo-...}.

\end{acknowledgments}

\begin{figure}[t!]
\includegraphics[scale=.29]{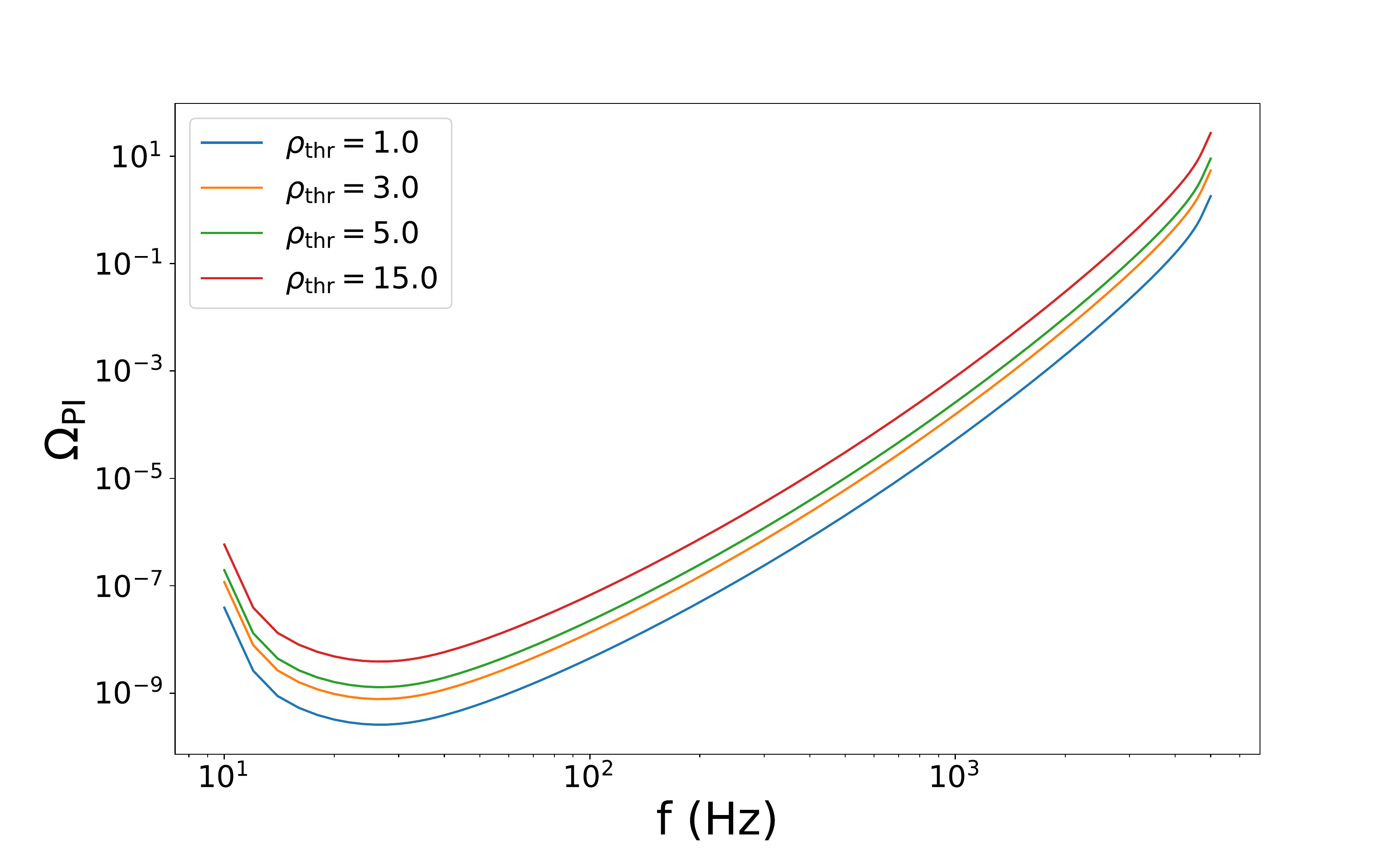}
\caption{Example of a power-law integrated sensitivity curve for the LVK detectors using O4 sensitivity and assuming a threshold SNR $\rho_{\rm thr}= 1, 3, 5, 15$ and an observation time $T=1$ year.}
\label{fig: PI curve}
\end{figure}

\appendix
\section{Power-law integrated sensitivity curves}
\label{PI}
We summarise the construction of  power-law integrated curves \cite{Thrane_2013}. Consider the signal-to-noise ratio (SNR) for a GW background $\Omega_{\rm GW}$ after observing time $T$ by a detector network  $M$: 
\begin{equation}
    \rho=\sqrt{2T}\left[\int_{f_{\rm min}}^{f_{\rm max}}df\frac{\Omega_{\rm GW}^2(f)}{\Omega_{\rm eff}^2(f)}\right]^{1/2},
\end{equation}
where the effective energy density reads
\begin{equation}
    \Omega_{\rm eff}(f)=\frac{2\pi^2}{3H_0^2}f^3\left[\sum_{I=1}^{M}\sum_{J>I}^M\frac{\Gamma_{IJ}^2(f)}{P_{nI}(f)P_{nJ}(f)}\right]^{-1/2},
\end{equation}
with $\Gamma_{IJ}$ the overlap reduction function and $P_{nI}$ the noise power spectral density of detector $I$. Assuming a power-law spectrum $\Omega_{\rm GW}(f)=\Omega_\beta(f/f_{\rm ref})^\beta$, one can calculate the value of $\Omega_\beta$ such that some SNR threshold $\rho_{\rm thr}$ is reached:
\begin{equation}
    \Omega_\beta=\frac{\rho_{\rm thr}}{\sqrt{2T}}\left[\int_{f_{\rm min}}^{f_{\rm max}}df\frac{(f/f_{\rm ref})^{2\beta}}{\Omega_{\rm eff}^2(f)}\right]^{-1/2}.
\end{equation}
This procedure is repeated for a series of $\beta$ values, e.g. $\beta\in\{-10,-9,\dots,9,10\}$. The PI curve is given by:
\begin{equation}
    \Omega_{\rm PI}(f)=\max_{\beta}\left[\Omega_\beta\left(\frac{f}{f_{\rm ref}}\right)^\beta\right].
\end{equation}
By construction, any line on a log-scaled plot, corresponding to a power-law GW background, which is tangent to the PI curve, will have an integrated SNR equal to the chosen threshold value $\rho_{\rm thr}$. A curve that falls below the PI curve would be observed with and SNR lower than $\rho_{\rm thr}$, whereas an SNR larger than $\rho_{\rm thr}$ is expected for curves that fall above the PI curve. The result of the above procedure is illustrated in Fig.~\ref{fig: PI curve} where the PI curve is shown for the LVK detectors at O4 sensitivity using a threshold SNR of $\rho_{\rm thr}=1, 3, 5, 15$ and assuming an observation time $T=1$ year.

\bibliography{FOPTbib}

\end{document}